\documentclass[conference]{IEEEtran}

\usepackage[utf8]{inputenc}
\usepackage[british]{babel}

\usepackage{booktabs,xspace,soul,csquotes,subcaption,microtype,graphicx,amsmath,amssymb}
\usepackage[nohyphen]{underscore}
\usepackage[style=ieee,mincitenames=1,maxcitenames=2]{biblatex}
\newcommand{\A}{\TextOrMath{\(A\)\xspace}{A}}
\newcommand{\B}{\TextOrMath{\(B\)\xspace}{B}}

\newcommand{\varc}{\TextOrMath{\(c\)\xspace}{c}}

\newcommand{\NA}{\TextOrMath{\(N_\A\)\xspace}{N_\A}}
\newcommand{\NB}{\TextOrMath{\(N_\B\)\xspace}{N_\B}}
\newcommand{\mID}{\TextOrMath{\(\mathit{mID}\)\xspace}{\mathit{mID}}}
\newcommand{\sID}{\TextOrMath{\(\mathit{sID}\)\xspace}{\mathit{sID}}}
\newcommand{\KAB}{\TextOrMath{\(K_{\A\B}\)\xspace}{K_{\A\B}}}
\newcommand{\KC}{\TextOrMath{\(K_\varc\)\xspace}{K_\varc}}
\newcommand{\KDF}{\texttt{KDF}\xspace}
\newcommand{\MAC}{\texttt{MAC}\xspace}
\newcommand{\Hash}{\texttt{H}\xspace}
\newcommand{\AEAD}{\texttt{AEAD}\xspace}
\newcommand{\concat}{\|\allowbreak}
\newcommand{\bits}{\{0,1\}}
\newcommand{\kappabits}{\bits^\kappa}
\newcommand{\starbits}{\bits^\ast}
\newcommand{\inkappabits}[1]{\TextOrMath{\(#1\in\kappabits\)}{#1\in\kappabits}}

\usepackage{hyperref,cleveref}
\hypersetup{hidelinks}

\begin{document}

\title{Securing Fleets of Consumer Drones at Low Cost}

\author{
  \IEEEauthorblockN{Nick Frymann and Mark Manulis}
  \IEEEauthorblockA{Surrey Centre for Cyber Security \\ University of Surrey \\ 
    \texttt{\{\href{mailto:n.frymann@surrey.ac.uk}{n.frymann},\href{mailto:m.manulis@surrey.ac.uk}{m.manulis}\}@surrey.ac.uk}}
}

\maketitle

\begin{abstract}

In recent years, the use and suitability of drones for many applications, including surveillance, search and rescue, research, agriculture and civil engineering, has greatly increased due to their improved affordability and improved functionality. However, low-cost consumer drones are rarely designed to work in fleets, which limits the applications for which business, research and individuals may deploy such drones. Proprietary, commercial and bespoke options are available at higher cost and existing solutions providing fleet functionality have limited security, if any, which excludes their use for sensitive applications. In this paper, we discuss the repurposing of consumer off-the-shelf (COTS) drones for use in secured fleets and provide the design, implementation and evaluation of a complete approach for creating end-to-end secured fleets of consumer drones (SFCD).

We present a protocol for securing communications within fleets whilst employing more efficient symmetric key cryptography throughout to reduce the impact of our security on the limited and resource-constrained COTS drones---exploiting the characteristics of a fleet with an online and central ground control station, which may act as a key distribution centre. The protocol allows an arbitrary number of channels to be established to authenticate and optionally encrypt real-time data transmitted on these channels. We also discuss routing in fleets, as well as the control and monitoring of them, to allow SFCD to be fully deployed---providing an extensive and thorough solution. Our experimental evaluation confirms the suitability of low-cost consumer drones for use in SFCD, with flight time impacted by only 9.9\% and worst-case bandwidth of 4.7Mibit/s.

\end{abstract}

\thispagestyle{plain}
\pagestyle{plain}

\section{Introduction}\label{sec:intro}

The market for consumer drones, or Unmanned Aerial Vehicles (UAVs), is booming; facilitated by the continual improvement in their hardware and software---and ever-decreasing prices. Major vendors such as DJI and Parrot are offering high-quality consumer drones for as low as 300 USD, equipped with high-quality video cameras, powerful processors, batteries supporting flights of up to 30 minutes, and a range of sensors for safe flight operations (e.g. collision avoidance, obstacle detection). The current consumer market is focused on providing easy-to-use, out-of-the-box command and control functionality for safe operation of \emph{individual} drones, controlled manually using R/C controllers or using smartphone applications, such as Parrot's FreeFlightPro\footnote{\url{https://www.parrot.com/us/freeflight-pro}}, connected to the drones via Wi-Fi. The latter offer more sophisticated means of semi-automated control using flightpaths and way points based on GPS coordinates.

In contrast, the majority of current and future business applications, in particular involving surveillance/monitoring in search and rescue operations, land surveying, ecological research, agriculture and civil engineering, require command and control systems that are able to support \emph{multiple} UAVs operating in a centrally-managed fleet, i.e. controlled by some operator from the Ground Control Station (GCS). Such functionality is currently not widely available on consumer drones; hence businesses are forced to resort to commercial providers offering bespoke and more expensive fleet management systems. The restrictive costs of commercial solutions and the current limitations on the use of consumer drones in fleets makes it hard for businesses to adopt drone technologies.

To this end, extending the functionality of low-cost consumer drones to support fleet operations would be an attractive proposition---especially if these changes can be limited to the software only. A number of open source projects, such as PaparazziUAV\footnote{\url{http://wiki.paparazziuav.org/wiki/Main\_Page}} and ArduCopter\footnote{\url{http://ardupilot.org/copter/}}, have been developed to provide extendable command and control functionality for enthusiasts who build their own drones. Most of these projects also support various consumer off-the-shelf drones by replacing their native software. However, the required support for fleet operations in these projects is limited: PaparazziUAV offers support for multiple drones, but uses its own communication protocol to control the drones---restricting users to just the PaparazziUAV ecosystem. ArduCopter uses a more popular communication protocol, MAVLink\footnote{\url{https://mavlink.io/en/guide/serialization.html}}, whose specification supports multiple drones based on unique system identifiers, allowing for a wide range of MAVLink-compatible GCS software to be employed. Despite this, GCS software available is often not designed with fleet-centric control in mind due to the single-drone, single-controller approach taken for many consumer drones currently on offer.

Besides, these projects provide only limited support for securing communications between two drones or a drone and its GCS. For instance, S-PPRZLINK\footnote{\url{http://wiki.paparazziuav.org/wiki/Pprzlink}} secures the execution of the PaparazziUAV's communication protocol, PPRZLINK, between two entities using public key cryptography. This approach would not scale well in a centrally-managed fleet environment. In academic literature, several proposals focused on securing communications between drones without considering the wider aspects of a fleet management system such as networking and mission control. For example, \citeauthor{won2015secure} \cite{won2015secure} designed a new key management protocol for UAVs supporting key agreement, non-repudiation, and revocation based on certificate-less signcryption and key encapsulation. \citeauthor{zouhri2016new} \cite{zouhri2016new} proposed an architecture for secure key management, access control and data transfer between the entities of a fleet, yet without actual protocols to instantiate the architecture. \citeauthor{blazy2017efficient} \cite{blazy2017efficient} used streams of one-time keys derived from a shared key to protect messages exchanged between a drone and the GCS. \citeauthor{maxa2015secure} \cite{maxa2015secure} focused on highly dynamic UAV ad-hoc networks and explored the use of various routing protocols (including AODV, OLSR and DSR) and their security in this context. We observe that typical consumer drones do not provide native support for these routing protocols; out of the box, these drones are typically configured to serve as IEEE 802.11 access points (APs) in master mode.

At the moment there is no complete system that addresses the different aspects of building secure fleets of drones and proposes solutions for managing the fleet and its missions.

\paragraph*{Our contribution} In this work we design, implement and evaluate a complete system, called SFCD, for secure management of fleets built from low-cost off-the-shelf consumer drones. Our SFCD system does not require any hardware modifications to the drones and comes with GCS software that provides centralised fleet-centric control for missions involving up to tens of consumer drones.

Each drone in the fleet is configured with an independent (symmetric) cryptographic key that it shares with the GCS. For each mission, the GCS and each participating drone establish several independent end-to-end secure channels, which are used to transmit different types of messages, including dedicated channels for control messages, telemetry data, and application payload data (e.g. video streams). The underlying SFCD handshake protocol uses provably secure lightweight mutual authentication protocol from \cite{Bellare1994} by \citeauthor{Bellare1994}, extended with derivation of channel-specific session keys. The actual transmission of messages over these channels is secured with the state-of-the-art authenticated encryption scheme from \cite{RFC7905} that supports authentication of additional data. In our system, the GCS can also be used to enable secure communication between any two drones participating in a mission. For this purpose, the GCS acts as a key distribution centre and securely transmits a fresh shared key to these drones---enabling them to execute the SFCD handshake and set up their own independent secure channels if required.

In terms of networking, our SFCD system establishes an IEEE 802.11 ad-hoc network between the drones and the GCS. The lack of support for routing protocols for ad-hoc networking in consumer drones is compensated by the deployment of efficient application-layer routing, in which the GCS dynamically sets the required routes to ensure that each drone maintains its connectivity to the GCS. Packets sent along multi-hop paths are additionally authenticated by a path key, distributed by the GCS which acts as a key distribution centre.

Our SFCD system uses its own addressing system for the drones and a dedicated application-layer protocol header that has been designed to establish, protect, and maintain all communication channels. This header is seen as `additional data' and is protected by the underlying channel security mechanisms.

The designed SFCD system has been implemented in a modular way, in which a standalone program runs alongside other software on the drone, such as PaparazziUAV, to provide secure communications according to our protocol and has been practically evaluated using the popular Parrot Bebop 2\footnote{\url{https://www.parrot.com/us/drones/parrot-bebop-2}} drone. Our experiments attest to the practicality of our SFCD system and its suitability for monitoring/surveillance applications. The conducted experiments show that our SFCD solution with its security protocols has only minor impact on the battery power consumption, leading to a 9.9\% reduction in flight time in comparison to the original PaparazziUAV software running on its own, and achieves throughputs of at least 4.7Mibit/s over distances of up to 260m per drone---capable of supporting real-time video streaming at the default quality and frame rate.

\paragraph*{Organisation}
The rest of the paper is organised as follows. In \Cref{sec:fleets}, we describe the main architecture, components and requirements of a general system for the centralised control of a Fleet of Drones (FoD). In \Cref{sec:engineering}, we introduce the design of our SFCD system, focusing on the engineering aspects behind software modification for consumer drones, using the example of Parrot Bebop 2 drones, and the realisation of all system components, including provision of security and routing mechanisms. This section contains specifications of the SFCD handshake, channel establishment protocols, application-layer routing and definition of the packet structure used in all SFCD communications. \Cref{sec:evaluation} experimentally evaluates the performance of the SFCD system, including its impact on battery power consumption and quality of communication links over distances, and discusses the suitability of the system for monitoring/surveillance applications requiring real-time data streaming.

\section{Fleets of Drones---System Architecture and Requirements}\label{sec:fleets}

Our work focuses on fleets of UAVs. Based on the classification by \citeauthor{8101984} in \cite{8101984}, for a Fleet of Drones (FoD), we consider up to tens of UAVs acting on instruction from a central operator, a GCS, with human oversight and some limited autonomy---such as flying to specific GPS coordinates. This differs from a drone swarm, which is a group of fully-autonomous UAVs employing swarm intelligence with little to no human oversight. Our work is centred on such fleets due to their potential use for current and future business applications.

In the following section we discuss the main characteristics, architecture and requirements for FoDs, including their control systems and communications.

\subsection{FoD Components}

\begin{figure*}
  \centering
  \captionsetup{justification=centering}
  \includegraphics[width=.75\linewidth]{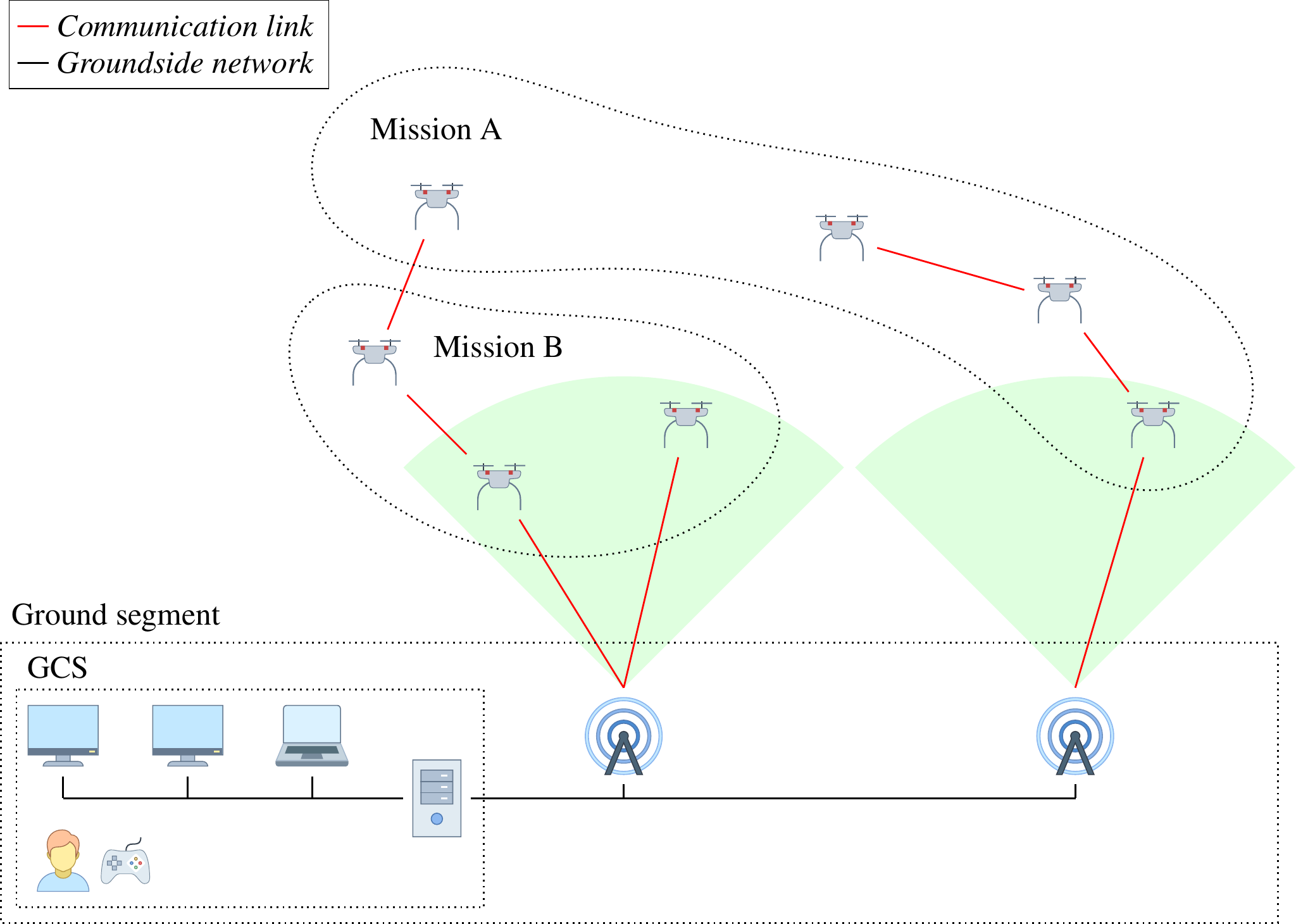}
  \caption{Example of a UAV fleet\protect\footnotemark}\label{fig:fod}
\end{figure*}

An FoD command and control system consists of the following components which work together to form a fleet architecture, exemplified in \Cref{fig:fod}.

\paragraph*{GCS} In an FoD, the GCS is the logical central controller for all UAVs in the fleet. The GCS has situational awareness regarding all actions performed by a fleet, including the status, location and connectivity of its UAVs. A fleet GCS is designed to provide seamless secure command and monitoring of multiple UAVs simultaneously.

\footnotetext{Iconography: \url{icons8.com}}

Typical GCS software allows UAVs to be programmed with flight plans and controlled from a control station. Specifically, these programs provide a GUI that allows the operator to:

\begin{itemize}
  \item view real-time UAV data (telemetry), such as battery, position, and orientation;
  \item instruct UAVs to fly to a single or set of GPS coordinate(s)---forming a flightpath; and,
  \item control UAVs manually using a joystick or game controller.\footnote{In a traditional consumer UAV system, manual control is often provided through a groundside R/C safety link.}
\end{itemize}

GCS hardware consists of a computer or computers running GCS software; operators use this to control and monitor the fleet. In \Cref{fig:fod}, the example GCS uses a gateway machine which acts as the demarcation point between the GCS and the rest of the groundside network. This gateway machine may run GCS server software; allowing multiple computers, used by fleet operators, to connect using GCS client software. Alternative approaches may incorporate multiple servers to manage fleet management workstations---or simply a single machine, such as a laptop, running both the operator's interface and integrated GCS software.

The PaparazziUAV project, an open-source ecosystem of UAV and GCS software, produces software which provides a messaging protocol, graphical interfaces for operators, a datalink program to manage communication channels and server software for handling communications between GCS instances. QGroundControl\footnote{\url{http://qgroundcontrol.com/}} and APM Planner 2\footnote{\url{http://ardupilot.org/planner2/}} are examples of GCS software which support the MAVLink UAV and GCS communications protocol.

\paragraph*{UAVs} Typical drone hardware consists of a number of components, such as inertial measurement units (IMUs), GPS receivers, cameras, motors, batteries, wireless radios, optical-flow sensors, ultrasound sensors and pressure sensors---as well as the airframe and propellers.

UAV software often consists primarily of autopilot software which manages flight operation, including:

\begin{itemize}
  \item monitoring and acting on sensor data by adjusting motor output or current action;
  \item reporting status to GCS (telemetry);
  \item flying to specified GPS coordinate(s) and maintaining position;
  \item returning to specified location when conditions are met (e.g. returning to launch after elapsed time); and,
  \item interpreting roll, pitch, yaw and throttle values from joystick input on GCS, or R/C controller, and altering motor output accordingly.
\end{itemize}

PaparazziUAV provides autopilot software for use with the rest of their ecosystem, which is supported on a number of platforms. ArduPilot is a MAVLink-compatible autopilot, suitable for use with MAVLink GCS software.

\paragraph*{Ground network and FoD communications}

The ground network uses communication hardware to create the groundside and ground-to-air networks. Network infrastructure may be used as part of the GCS when multiple machines are used to manage the fleet. In \Cref{fig:fod}, two wireless APs are used to form the ground-to-air network and a wired network for the groundside network. This includes the GCS which uses the groundside network to allow communication between operator machines and the gateway server. Alternatively, a single machine with a wireless network interface card (NIC) may form the entire GCS and groundside network.

The GCS uses the groundside network for groundside component communication and the ground-to-air network for communicating with UAVs through telemetry (downlink) and uplink channels, provided by the relevant wireless hardware (e.g., IEEE 802.11 APs).

Fleet UAVs form a mesh (ad-hoc) network to facilitate communication over a larger geographical area. In \Cref{fig:fod}, UAVs not in range of the groundside APs are served by other UAVs in range of the existing network via multi-hop communication links.

Standard communication channels formed between the GCS and each UAV include downlink (telemetry---UAV status), uplink/datalink (control), and other channels such as video or sensor data streams. To send commands to UAVs and receive status updates, protocols such as MAVLink, PaparazziUAV's PPRZLINK or proprietary protocols are used. These protocols are usually not protected---a notable exception is the S-PPRZLINK, a security add-on for PPRZLINK, which uses elliptic curve cryptography, namely Curve25519 \cite{Bernstein2006} key exchange, Ed25519 \cite{Bernstein2012} digital signature, and  ChaCha20-Poly1305 \cite{RFC7539} authenticated encryption scheme. Instead of S-PPRZLINK, we use our approach presented in \Cref{sec:protocol}---which uses preshared keys and the GCS as a key distribution centre---due to its simplicity, superior efficiency and better suitability for FoDs given the low number of drones and available connectivity with the GCS.

\subsection{Mission Phases}

A typical FoD mission consists of four main phases. The fleet may be engaged in multiple simultaneous missions; in \Cref{fig:fod}, we show two missions running concurrently. Missions do not require their UAVs to all be in the same phase at any given time.

\paragraph{Configuration} Networking, communication channels, software and keys are configured on the GCS and UAVs. Autopilot software is installed on UAVs if necessary. This phase should only be completed once for each component of the fleet, after which these components may be reused for many future missions.

\paragraph{Initialisation} The GCS readies the UAVs for the mission---preloading UAVs with flightpaths, roles and other instructions. The GCS can dynamically initialise UAVs at any time during a mission---they need not all be initialised simultaneously.

\paragraph{Execution} UAVs perform mission objectives, such as flying to a specified location and collecting data, such as video. The operator oversees and controls the fleet from the GCS.

\paragraph{Completion} The mission is completed when all objectives have been met. Some missions may be ongoing, running indefinitely, and may never finish---such as continuous surveillance patrols. UAVs may be expected to fly to specified GPS coordinates upon the completion of a mission or before their battery is fully depleted.

\subsection{FoD requirements}

In order to be deployable for various applications, FoD systems must support the following five overarching goals:

\paragraph{Control} For operational and safety purposes, FoD systems must come with an appropriate GCS to provide fleet operators with situational awareness. The corresponding GCS must at least be able to simultaneously track and control multiple fleet UAVs throughout the different mission phases and present real-time status information about individual fleet UAVs to facilitate operator's decision-making.

\paragraph{Infrastructure} FoD systems must be able to use provided networking infrastructure for the purpose of communication. Existing infrastructure, such as the wireless APs on buildings (e.g. for a perimeter patrol), may be exploited to provide ground-to-air communications. The GCS may require additional internal infrastructure for serving multiple workstations for operators to use.

\paragraph{Communication} UAVs in a fleet must be able to communicate with one another and their GCS. UAVs should forward and route messages for other UAVs and the GCS through multi-hop communication links, as in \Cref{fig:fod}, to provide better geographical coverage.

\paragraph{Security} To secure fleet operations and to allow for the use of FoDs for sensitive applications (e.g. surveillance), all communications within the fleet, including control messages and application data must be protected. In particular, we require that all end-to-end communications between the GCS and UAVs, as well as between individual UAVs if needed, are mutually authenticated and encrypted to an appropriate standard where required---with consideration for the possible impact on UAV performance. All communication channels must therefore be authenticated, with encryption used for sensitive channels such as video, status (e.g. current location) and high-level control. Some channels may not require encryption, such as low-level joystick control\footnote{The current location of the UAV would need to be known as well as localised conditions such as wind speed and direction to make use of this.} (e.g. roll, pitch, yaw and throttle values).

We assume that UAVs cannot be easily compromised whilst airborne. In particular, in order to access session-specific key material the attacker would need to ground and physically access the drone, which effectively rules out real-time compromise attacks on the drones participating in the mission. Despite this, we still require that the compromising of a UAV does not weaken end-to-end channel security for other UAVs in the fleet and that the GCS must attempt to find alternative routing paths upon discovering connectivity issues---an extension of the communication requirement.

\paragraph{Performance} For an FoD to be fit-for-purpose, it must provide sufficient performance, quality of service and reliability for its applications. In particular, the adopted security mechanisms should be as lightweight as possible so as to have minimal impact on the battery power and to allow for sufficient throughput of payload data (e.g. video). Sensitive control messages should be delivered with minimal delay, possibly using appropriate traffic prioritisation mechanisms.

\section{Secured Fleets of Consumer Drones---Our Engineering Approach}\label{sec:engineering}

In this section, we show how appropriate FoD systems can be engineered at low cost from traditional COTS drones. Our solution for SFCD does not require any hardware modifications and, despite using a particular drone model to demonstrate the system, the underlying steps and developed building blocks are platform-agnostic and can be adopted for many other low-cost COTS drones. Our approach is designed to minimise changes to the operating system, kernel and other software components of consumer drones. This is due to the fact that some platforms may not readily support the changes required for communicating with other drones or the GCS.

 \begin{figure}
  \centering
  \includegraphics[width=\linewidth]{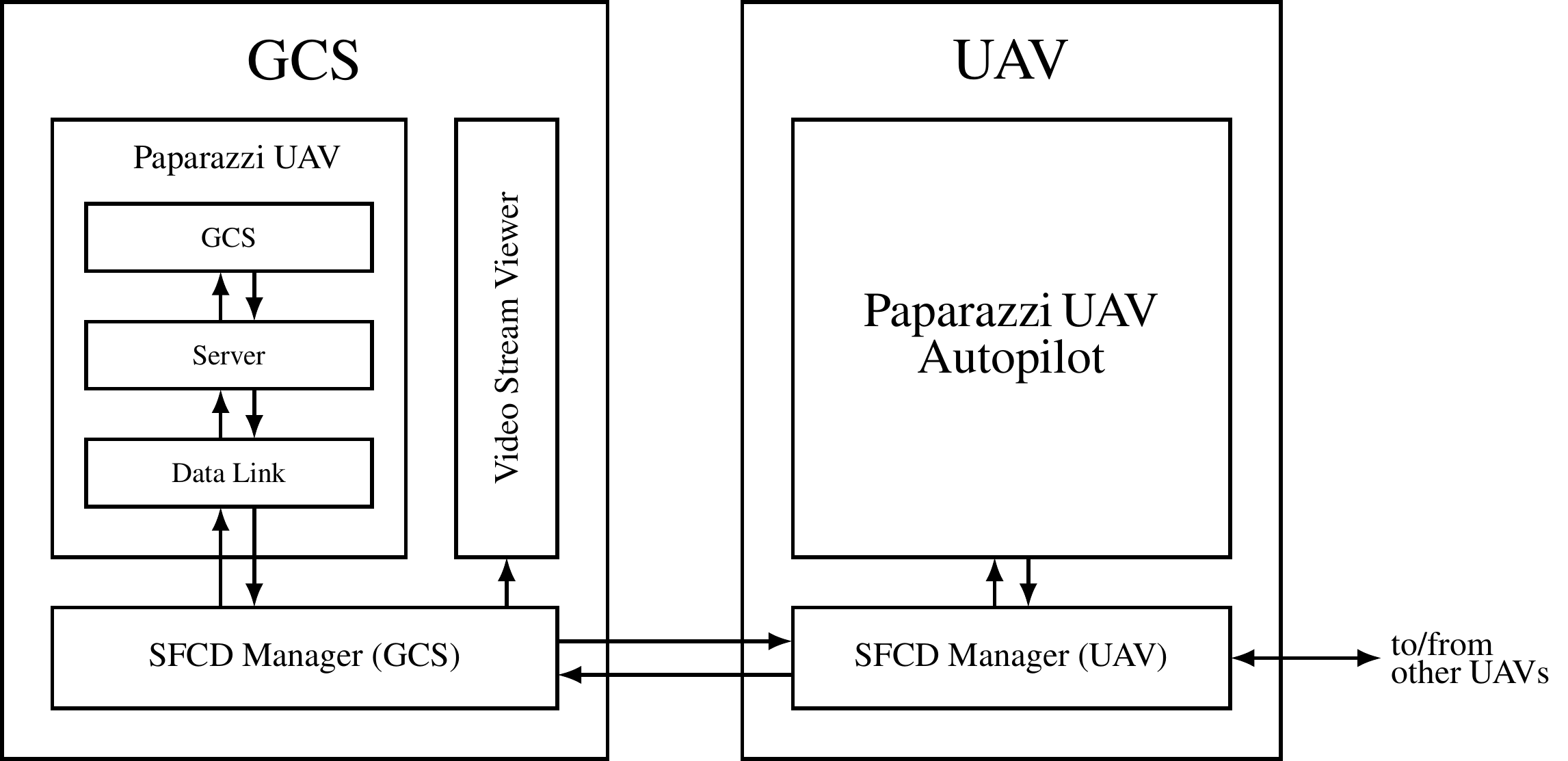}
  \caption{System architecture for our SFCD approach}\label{fig:block}
\end{figure}

\subsection{Consumer drones suitable for SFCD}

Typical low-cost COTS drones supporting video capture and transmission over ranges of about 200 to 300 metres, with flight times around 20 to 30 minutes, can be found from 150 to 400 USD. They come with on-board sensors such as gyroscopes, barometers, accelerometers, GPS, and magnetometers, with measurements helping to operate and position the drone. These drones can typically be controlled by a smartphone over Wi-Fi or through a dedicated R/C controller.

In our solution, we use the popular Parrot Bebop 2 drone. It is a quadcopter equipped with a dual-core Parrot P7 Cortex A9 CPU, 8GB onboard flash storage, 3350mAh battery---providing 30 minutes of flight time---and support for IEEE's 802.11a/b/g/n/ac networking standards through two aerials (2.4 and 5GHz) with a signal range of 300m. At the moment of writing, the Parrot Bebop 2 can be purchased for less than 350 USD. This drone is comparable with other drones in the same price segment such as the DJI Spark\footnote{\url{https://www.dji.com/uk/spark}}, Yuneec Mantis Q\footnote{\url{https://us.yuneec.com/mantis-q}} and Ryze Tello\footnote{\url{https://www.ryzerobotics.com/tello}}.

The Parrot Bebop 2 runs version 3.4.11 of the Linux kernel and BusyBox\footnote{\url{https://www.busybox.net/}} v1.25.0. The \texttt{/sbin/broadcom\_setup.sh} script is executed on startup and configures the network adapter, creating the drone's Wi-Fi AP---to which a compatible controller, such as a smartphone running Parrot's FreeFlight Pro application or a Parrot SkyController\footnote{\url{https://www.parrot.com/uk/support/products/parrot-skycontroller}}, may connect. A DHCP server is also started.

\subsection{Configuration changes to network and onboard software}

In order to adapt COTS drones for FoD use, some preliminary changes in configuration will be needed in most cases. In particular, the original networking setup will require alteration. This is because, out of the box, COTS drones are often configured as wireless APs allowing a single controller to connect and control the drone. COTS drones will not be able to communicate in a fleet as required with this configuration, i.e. establish simultaneous connections with the GCS and other drones in a peer-to-peer manner.

In addition to networking, many COTS drones are equipped with closed-source autopilot and GCS software (e.g. smartphone flight applications) which cannot be directly modified for FoD mission control purposes. It is therefore necessary to replace them with appropriate software that allows such modifications. Fortunately, the aforementioned open-source projects, PaparazziUAV and ArduCopter, can be used on various COTS drones and replace the original manufacturer's software whilst remaining compatible with the drone's hardware.

In our approach, these steps have been taken for the Parrot Bebop 2 drones. The original wireless network adapter (Broadcom BCM43526 802.11a/b/g/n/ac Wireless Adapter) was configured to support the ad-hoc mode such that the GCS and all fleet drones can connect to the same ad-hoc network upon boot. In addition, Parrot's closed-source autopilot software, \texttt{dragon-prog}, was replaced with the PaparazziUAV autopilot software, version  5.14.0. The main reason for choosing PaparazziUAV's autopilot is that its counterpart GCS software, unlike other projects, already provides a graphical interface with status display and controls for multiple drones simultaneously, a key requirement for FoD mission control.

\paragraph*{Compatibility}

Our SFCD solution requires that UAVs participating in the fleet use standard IP networking over an IEEE 802.11 ad-hoc network. Given that IEEE 802.11 is the de-facto communication standard for many COTS drone manufacturers, our solution becomes compatible with a wide range of consumer drones that support Wi-Fi. PaparazziUAV supports STM32- and LPC21-series microcontrollers, used by most autopilot boards, as well as Pixhawk and Parrot's AR.Drone 2.0, Bebop, Bebop 2 and Rolling Spider\footnote{\url{http://wiki.paparazziuav.org/wiki/Category:Autopilots}}.

Strictly speaking, UAVs are not required to run PaparazziUAV's autopilot software. It is possible to use alternative autopilot software (e.g. ArduCopter) on the drones, but doing so may require alteration to compatible GCS software (e.g. QGroundControl) to provide a more fleet-centric GUI for the monitoring and control of fleets. We use PaparazziUAV to test our solution as a suitable fleet interface is already implemented.

We implement a standalone program, the SFCD manager (see \Cref{fig:block}), that handles the handshake and the forwarding, authentication and encryption of packets. It intercepts these packets from either the GCS or autopilot software and applies our security protocol---without any changes needing to be made to the original software on any fleet participant.

\subsection{SFCD multi-hop connectivity and routing}\label{sec:multihop}

UAVs participating in an SFCD can communicate with other fleet UAVs and the GCS. Considering UAVs may not be in the direct range of the GCS or ground infrastructure, a multi-hop connection through other drones needs to be established where necessary with appropriate mechanisms for packet routing.

There exists numerous routing protocols that may be suited to a Flying Ad-hoc Network (FANET). \citeauthor{BEKMEZCI20131254} \cite{BEKMEZCI20131254} discuss the features of a FANET when compared to Mobile Ad-hoc Networks (MANETs) and Vehicular Ad-hoc Networks (VANETs), with one of the primary distinctions being the mobility degree of nodes in a FANET---which affects the design and effectiveness of protocols. \citeauthor{7069210} \cite{7069210} present the results of experiments performed using P-OLSR (Predictive Optimised Link-State Routing) for use in FANETs. The authors implemented P-OLSR based on an existing OLSR implementation. \citeauthor{OUBBATI201729} \cite{OUBBATI201729} discuss a number of approaches to routing in a FANET and the classification of these approaches, including location-based routing. AODV, DSDV and OLSR were experimentally analysed by \citeauthor{7226085} \cite{7226085}, with findings suggesting that OLSR outperforms AODV and DSDV.

Despite this, given that our approach aims to be as platform-agnostic as possible, many routing protocols cannot be practically used due to limited platform support of their implementations, if any such implementations exist. With the situational awareness of the GCS, including the current and planned locations and paths of its UAVs and the relatively small size of our fleets, the additional network and computational overhead---and implementation time---of a fully-dynamic and decentralised routing algorithms is \emph{not} outweighed by its advantages. Location-based and predictive routing may be offloaded to the GCS since it has knowledge of the network's current and planned future physical topology, which avoids unnecessary computation on UAVs.

We observe that the use of multi-hop routing may not be completely avoided since this mechanism allows the fleet to extend its geographic coverage whilst maintaining connectivity with the GCS. However, the use of multi-hop routing should be minimised due to the communication bottleneck and the increased usage of already limited resources (e.g. CPU, battery power) on consumer UAVs that will be forwarding packets.

Since the GCS has situational awareness about the mission, including the planned and current location of all drones, coupled with the assumption that a typical SFCD will consist of up to few tens of UAVs, it is more practical to avoid the use of costly ad-hoc routing protocols and adopt an approach where the corresponding routes are either statically configured or set dynamically by the GCS and changed on-demand, as part of the operational mission control.

In our SFCD approach, we adopt \emph{application-level routing} where the GCS can set multi-hop routes dynamically so that each UAV not in direct range of the ground network learns the next-hop UAV which is responsible for forwarding its packets. A dedicated routing tag field is reserved in our SFCD packet format (see \Cref{sec:packet}) for packets that require multi-hop routing. 

In order to protect multi-hop routing from an outsider who may attempt to flood the network with packets, the GCS can securely distribute a fresh \emph{path key} to all UAVs on a given path when it sets or updates routes. This path key will be used by the sending UAV to authenticate its packets and by every in-path UAV to check validity prior to forwarding the packet. We stress that path keys are used on top of end-to-end protection and so an outsider in possession of a path key would only be able to flood that path with its own packets but not to compromise the end-to-end security between the fleet UAVs and the GCS. Besides, as already mentioned, the attacker would need to ground and physically access the UAV in order to obtain the current path key from the drone's volatile memory. In this case GCS will no longer receive the expected regular status messages from the affected UAVs and can take measures to remove them from the ongoing mission and update the affected path key for the remaining UAVs. After a timeout specified in mission parameters, GCS can attempt to discover an alternative route to the remaining UAVs. If unsuccessful, UAVs may be configured to fly to specified coordinates or their launch location after not receiving control messages from the GCS for a specified duration.

\subsection{SFCD communication channels}\label{sec:data}

For PaparazziUAV's autopilot and GCS software, UDP is used to send commands to UAVs and to receive regular status updates in the PPRZLINK format. Status updates are on average 55--108 bytes, with control around 50--68 bytes. Low-level joystick values sent to the UAV are 14 bytes long. These channels form the unidirectional telemetry (status) and datalink (control) channels between the GCS and UAVs. With our packet format in \Cref{sec:packet}, some unavoidable overhead is introduced in authenticating and securing these channels.

Enabling video streaming for both the bottom and front-facing cameras on the Parrot Bebop 2 drone, with PaparazziUAV's default quality settings, results in RTP packets of size 298--342 bytes at 5 FPS and 1054--1462 bytes at 5 FPS, respectively. Scaling, compression and FPS parameters may be adjusted to give improved quality or latency. This forms another unidirectional communication channel between a video stream viewer on the GCS and the video transmission software on the UAV. UAVs equipped with other sensors of use for a given use case may be configured to stream readings on other data stream channels.

\subsection{Securing SFCD communication channels}

The aforementioned communication channels must be sufficiently secured to allow the use of SFCD in sensitive applications. The adopted security mechanisms need to be lightweight in order to have minimal impact on the limited resources of consumer drones.

Given the online presence of the GCS for the entire SFCD mission duration and the relatively small number of participating UAVs, we base our security approach on the efficient on-demand key distribution technique using symmetric cryptography. In particular, we consider GCS as a key distribution centre with individually preshared independent master keys between the GCS and each UAV. Upon initialisation of each drone, a lightweight handshake protocol will be executed to establish fresh session keys for each communication channel needed between the GCS and the UAV. If at some point the application requires two drones to have end-to-end secure communication, both drones will receive a fresh shared key from the GCS over their established secure channels and run the same handshake protocol with each other to establish their own secure channel(s). Similarly, to protect application-layer routing against flooding attacks as discussed in \Cref{sec:multihop}, if the GCS dynamically sets a multi-hop route for a new UAV, then this UAV and all UAVs on its path to the GCS will receive a fresh path key from the GCS over the established secure channels.

\subsubsection{SFCD Handshake}\label{sec:protocol}
Our SFCD handshake builds on the provably secure lightweight MAP1 protocol from \cite{Bellare1994}, enriched with information related to the identities of communicating parties (e.g. GCS, UAVs), derivation of various SFCD-specific channel keys, and additional public parameters such as mission and session IDs, and verified using Scyther\footnote{\url{https://people.cispa.io/cas.cremers/scyther/}}. As proven in \cite{Bellare1994} this protocol offers mutual authentication and secure session key establishment. The protocol is executed over TCP which provides reliability and correct ordering and is essential for the handshake to be completed successfully---UDP would require additional application-layer error and retransmission handling. 

The SFCD handshake protocol is executed between two parties, an initiator \A and responder \B. In the context of SFCD, for GCS-to-UAV communications, the initiator role will be taken by the GCS, whereas for UAV-to-UAV communications, any UAV can serve as the initiator. 

It is assumed that prior to the protocol execution both parties are initialised with a secure preshared key \inkappabits{\KAB} for a suitable security parameter \(\kappa\). Note that preshared keys for GCS-to-UAV communications may be uploaded securely on the UAVs over a wired USB connection during the setup procedure (e.g. during the changes to network and on-board software configurations), whereas for UAV-to-UAV they can be distributed on demand to both UAVs by the GCS over active end-to-end secure channels.

The SFCD handshake protocol uses standard cryptographic building blocks. An unforgeable message authentication code \(\MAC:\kappabits\times\starbits\mapsto\kappabits\) for mutual authentication of parties, a secure key derivation function \(\KDF:\kappabits\times\starbits\mapsto\kappabits\) for deriving session and channel-specific keys, and a cryptographic hash function \(\Hash:\starbits\mapsto\kappabits\) for computing the protocol transcript-dependent session ID. The mission ID, \mID, used in the protocol is assumed to be chosen by the GCS for the entire mission and sent to each UAV in that mission as part of the first protocol message. It will be used by all UAVs participating in the same mission upon execution of further handshake protocols for their UAV-to-UAV communications.

The SFCD handshake involves three communication rounds with final key derivation step as specified in the following:

\noindent\ul{Communication:}
\begin{enumerate}
  \item[1.] \A picks its nonce \inkappabits{\NA} and sends \((\A,\allowbreak\NA,\allowbreak\mID)\) to \B.
  \item[2.] \B picks its nonce \inkappabits{\NB}, computes \(\mu_\B = \MAC(\KAB, \B \concat \A \concat \NB \concat \NA \concat \mID)\), and sends \((\B,\NB,\mu_\B)\) to \A.
  \item[3.] If \(\mu_\B\) verifies, \A sends \(\mu_\A = \MAC(\KAB,\A \concat \B \concat \NA \concat \NB \concat \mID)\) to \B and proceeds with step 5.
  \item[4.] If \(\mu_\A\) verifies, \B proceeds with step 5.
\end{enumerate}
\noindent\ul{Key derivation:}
\begin{enumerate}
  \item[5.] Compute the session ID \(\sID = \Hash(\A \concat \B \concat \NA \concat \NB \concat \mID)\). For each channel ID \(\varc\), derive the corresponding channel key \(\KC = \KDF(\KAB,\A \concat \B \concat \NA \concat \NB \concat \mID \concat \varc)\).
\end{enumerate}

Note that a different constant \varc is used to derive each channel key \KC. It is assumed that both parties share knowledge about channels that they need to create.

In our implementation of SFCD handshake, we adopt 256-bit preshared master keys and use the very efficient BLAKE2 construction \cite{blake2} to instantiate \Hash and \KDF. Our \KDF produces 256-bit channel keys based on the preshared key \KAB. We use BLAKE2 and truncate its output to obtain the 8-bit \sID. Nonces \NA and \NB are 128 bits long, giving a combined entropy of 256 bits for the handshake. For the \MAC we adopt HMAC-SHA512-256. These instantiations are known to provide sufficient security for the cryptographic building blocks that they represent. Our SFCD code uses the popular libsodium cryptography library that provides the required implementations, with faster primitives\footnote{\url{https://download.libsodium.org/doc/}}, and can be easily ported to different platforms. In particular, its cross-compatibility makes it a good candidate for a SFCD, which may employ a variety of UAV models.

\subsubsection{Protecting SFCD channels} Each session-specific channel key \KC derived in the SFCD handshake protocol protects packets that will be exchanged over the corresponding channel \varc. For SFCD channel protection, we adopt the standard approach based on authenticated encryption with associated data (AEAD) \cite{bellare2003conventional}. In our SFCD construction we instantiate \AEAD using ChaCha20 stream cipher and Poly1305 for \MAC as specified in \cite{RFC7539} and implemented in the libsodium library. This combination provably offers 256-bit security \cite{procter2014security} and its efficiency is particularly well suited for SFCD communications. 

All end-to-end channels in SFCD require authentication, but payload encryption may be optional in some cases. The above AEAD approach allows both requirements to be achieved depending on the needs of the channel. Each SFCD packet header is treated as associated data and will be authenticated by default. For the packet data, authentication is always in place, with optional encryption, and is fixed for each channel in advance. 

\subsubsection{Structure of SFCD packets}\label{sec:packet}

Our SFCD approach proposes the use of a single packet format (as defined in \Cref{tab:encap}) for the protection of all data sent on any channel that is established following the handshake, in all GCS-to-UAV and UAV-to-UAV communications. This has advantage that the same protection mechanism and packet encapsulation method can be used across all channels and types of data. The packet format primarily acts as a wrapper to add  authentication and optional encryption for packets that would normally be transmitted without either.

\begin{table}
  \centering
  \caption{SFCD packet format}\label{tab:encap}
  \begin{tabular}{ll}
    \toprule
    Field      & Size     \\ \midrule
    Src ID     & 1 byte   \\
    Dest ID    & 1 byte   \\
    Mission ID & 1 byte  \\
    Type       & 1 byte   \\
    Session ID & 1 byte   \\
    Seq No     & 3 bytes  \\
    Payload    & ---      \\
    MAC tag    & 32 bytes \\ 
    Routing tag (multi-hop packets only) & 32 bytes \\ \bottomrule
  \end{tabular}
\end{table}

Our SFCD packet structure allows for 255 active drones in a single mission (one reserved for the GCS). It includes the mission ID, \mID, to distinguish between up to 256 missions that may be simultaneously taking place on the same network---requiring only one NIC as a minimum on the GCS. This allows the GCS to run multiple instances of its control software for different mission IDs if necessary---this may be on multiple workstations with individual operators, for example. Full bytes are used when identifying missions and UAVs to allow for a more straightforward implementation and future-proofing.

The type field indicates the payload type, or communication channel, such as video, status update (telemetry) or control (datalink). It may be used to prioritise routing, with up to 256 types possible. The type field is used to identify the channel key \KC required for processing the packet.

The session ID field, which contains the 8-bit \sID computed during the SFCD handshake protocol, will be checked by the recipient before it starts processing the received packet.

After completion of the handshake and establishment of the secure channels, the MAC tag will always contain the MAC value computed using \AEAD over the packet header and the payload field. The payload is encrypted for sensitive channels as defined in the mission parameters---usually expected to be video, status, and control channels.

We observe that our SFCD packet format adds 40 bytes of overhead, mainly due to the use of the MAC field, to all packets exchanged between GCS and UAVs that are in its direct range. Packets that must be routed via other UAVs in a multi-hop fashion have 72 bytes of overhead due to the use of the additional routing tag for verified routing.

The routing tag is needed only for packets that must be routed along some multi-hop path between the GCS and some UAV that is not in the direct range of the ground network (see \Cref{sec:multihop}). If present, the routing tag contains a MAC value computed by the sender using the path key distributed earlier by the GCS and known to all UAVs on the path. The receiving UAV will verify the tag before forwarding the packet.

\section{Experimental Performance Evaluation and Real-Time Streaming}\label{sec:evaluation}

\begin{figure*}[ht]
  \centering
  \begin{subfigure}[t]{\columnwidth}
    \includegraphics[width=\linewidth]{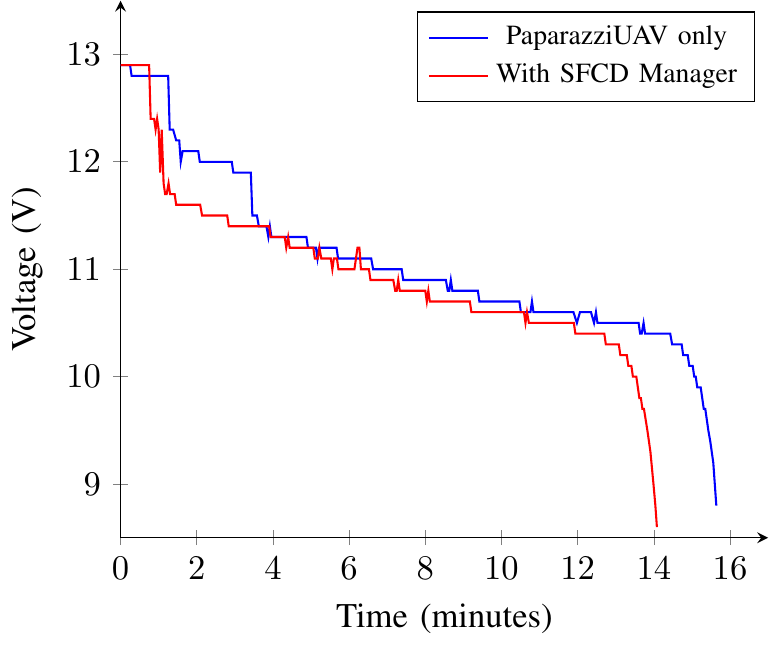}
    \subcaption{Power consumption comparison between PaparazziUAV on its own and PaparazziUAV with SFCD Manager running---with full throttle and front camera streaming at default quality}\label{fig:battery}
  \end{subfigure}\hfill%
  \begin{subfigure}[t]{\columnwidth}
    \includegraphics[width=\linewidth]{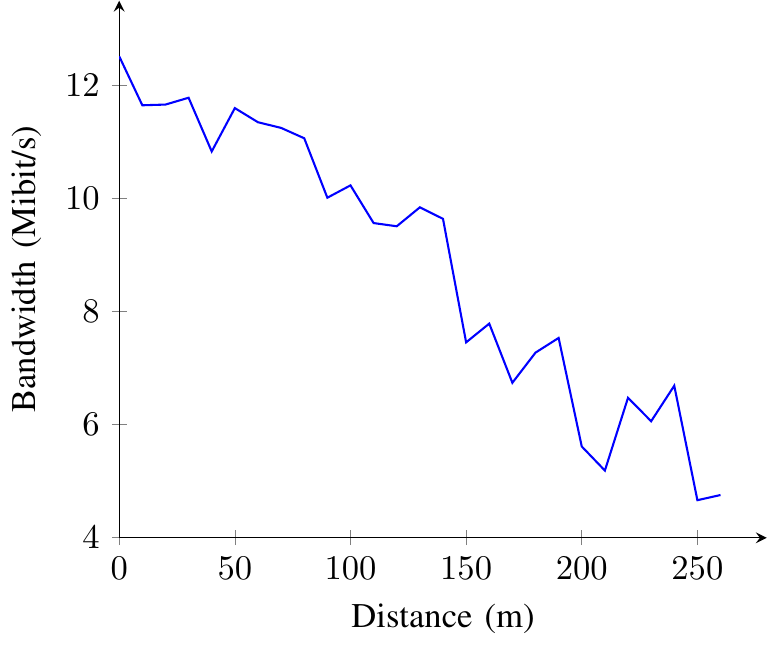}
    \subcaption{Drone-to-drone bandwidth measurements as distance increases}\label{fig:bwidth}
  \end{subfigure}
  \caption{SFCD with Parrot Bebop 2 drones---battery power consumption and bandwidth over distance}
\end{figure*}

In this section, we evaluate performance of our implemented SFCD solution in practice using Parrot Bebop 2 drones. Our experiments cover the impact of the implemented security mechanisms on the power consumption of the drones and their communication performance within the fleet and with the GCS. We use the obtained results to analyse suitability of our SFCD approach for FoD applications that require real-time monitoring/surveillance.

\subsection{Battery power consumption and flight time}\label{sec:eval:battery}
The impact of our SFCD approach on battery power is evaluated by comparing voltages over time for a Parrot Bebop 2 drone running PaparazziUAV software with and without our SFCD security mechanisms. We note that the manufacturer's flight time for the 3350mAh battery is claimed to be up to 30 minutes\footnote{\url{https://www.parrot.com/us/spareparts/drones/power-battery-bebop-2-power}} and that as soon as the battery output drops to 8--9V, the drone turns off.

Our power consumption experiments are based on the worst-case scenario where PaparazziUAV software on the Parrot Bebop 2 drone streams front-facing camera video at default quality whilst running the drone at \emph{full throttle}. The results presented in \Cref{fig:battery} show changes in power consumption measured on a single drone handling its own packets---that is, the verification and decryption of incoming packets and the encryption and signing of outgoing packets. We observe that our SFCD implementation impacts flight time by a maximum of 1.56 minutes, resulting in a 9.9\% reduction in comparison to the original flight time. Note that as voltages in \Cref{fig:battery} are measured for the drone at full throttle, the resulting running time is shorter than claimed by the manufacturer. The 9.9\% reduction equates to 27 minutes of normal flight, where throttle is not constantly maxed, based on the manufacturer's estimated flight time of 30 minutes.

As not all drones in the fleet would be required to stream video for surveillance applications, we expect that this maximal reduction in flight time would affect only drones that are tasked with streaming. We also note that drones tasked with forwarding packets on a multi-hop path would be required to verify routing tags, which would have similar impact on their power consumption compared to the verification of incoming packets addressed directly to the drone.

\subsection{Communication performance over distance}\label{sec:eval:comm}
The impact of our SFCD implementation on the bandwidth (data throughput) is estimated based on the measurements affecting drone-to-drone communication links only since the networking equipment of the GCS can be easily upgraded to be at least as powerful as the limited and non-upgradable COTS drone components. Our bandwidth measurements were obtained using  \texttt{iperf3}\footnote{\url{https://iperf.fr/}} on a communication link between two Parrot Bebop 2 drones secured with our SFCD solution. We note that the actual time for setting up a secure SFCD channel between the drone using SFCD handshake protocol (see \Cref{sec:protocol}) took 6--7ms on average to be completed.

The obtained bandwidth measurements over an SFCD-protected communication link are plotted in \Cref{fig:bwidth} in dependency of the distance between two drones. Tests were carried out every 10m with a drone distance from 0m to 260m---after which point, the \texttt{iperf3} client could not connect to the server and no packets were successfully returned with \texttt{nping}\footnote{\url{https://nmap.org/nping/}}. In our experiments, network latencies based on measured round-trip times with \texttt{nping} remained within 1 to 3ms, with all packets delivered and returned successfully.

As an outcome, the SFCD-protected communication links can support throughputs of up to 12.5Mibit/s. As the distance between the drones increases, the available bandwidth on their communication link decreases on average by 0.03Mibit/s/m. The maximum distance for an SFCD-protected communication link between two Parrot Bebop 2 drones can reach up to 260m---with line-of-sight and no other Wi-Fi channels in use nearby. Due to the rapid signal degradation from 260m onwards, no connection was observed between the two drones at 270m---before 260m, no packet loss was observed within the testable range.

The optimal topology for covering the largest geographical area is a star, where an AP for the GCS is centrally positioned, with drones 260m away from this AP concentrically. In order to cover the furthest distance, a line of drones at their maximal distance is required. The worst bandwidth measurement of 4.7Mibit/s, at the furthest distance of 260m, easily accommodates the data requirements of drones streaming video and telemetry at their default rate, as discussed in the following section.

\subsection{Impact on applications requiring real-time streaming}\label{sec:eval:stream}
In the following, we study the implications of our experimental SFCD performance evaluation in FoD applications related to monitoring/surveillance---where real-time data streaming from FoD to the GCS is of particular importance. Our analysis is based on video streaming at a resolution of \(288\times200\) at 5 FPS, which is PaparazziUAV's default for the front-facing camera on the Parrot Bebop 2. The lower bound of 4.7Mibit/s allows for a considerable increases in resolution if required. Streaming video at this quality, with a packet size of 1054--1462 bytes plus 40 bytes (or 72 for packets sent on multi-hop routes) for our packet format, gives an upper bound of \((5\cdot(1462+40))\div128=58.67\)Kibit/s (or 59.92 for multi-hop packets) for a single video stream. The observed bandwidth of a single connectivity link, 4.7--12.5Mibit/s, may handle \emph{theoretically} 82 such video streams.

As the number of nodes on a multi-hop path increases, so will the packet latency---provided the none of bandwidth constraints on the path are exceeded. The average round-trip time for small `ping' packets was 3ms, resulting in a 1.5ms penalty per hop for a unidirectional data stream. This gives an indication of the video or manual control latency that may be experienced for large fleets incorporating multi-hop packet forwarding. Additionally, the battery consumption of the drones providing such paths will be further impacted as in \Cref{fig:battery}. For each drone providing a communication link to others, an additional 9.9\% reduction in flight time is expected for every other drone it serves. A drone forwarding packets for the GCS and other drones may be required to throttle its processing of packets to ensure it can maintain steady flightstack processing if airborne.

Communication links within our SFCD may easily handle the default-quality video streams of many drones, with the worst bandwidth performance of any single link measured to be 4.7Mibit/s. However, the battery performance of drones providing mutli-hop connectivity will be impacted by around 9.9\% for each drone served. It is recommended that no more than three hops are present, or equivalently, no drone serves more than three others, as the flight time of the drone providing the connection to the GCS's AP will be adversely impacted. We suggest that our solution is well-suited to semi-automated surveillance or monitoring due to the reasonable bandwidths available, the limited performance of full manual control (e.g. joystick input), and the possibility to extend distances beyond 260m---up to 1040m if using three hops.

\section{Conclusion}

In this paper we described our engineering approach for building secure fleets based on low-cost off-the-shelf consumer drones. Our SFCD solution is modular and does not require extensive or intrusive changes to the consumer drones. We considered not only security in fleet communications but also issues around multi-hop routing, mission control and monitoring of fleet drones, offering a holistic approach to allow businesses, or even individual users, to adopt affordable small fleets of drones for their use cases.

Our SFCD system is designed with the limitations of consumer drones in mind, particularly their flight time performance, which results in a carefully-considered trade-off between security, network and communication performance. Consumer drones are more attractive due to their reduced cost, but inherently cannot provide the same performance as expensive commercial or bespoke solutions---a compromise between cost and quality. 

We exploit the online and central nature of the GCS to provide centralised routing management and key distribution, resulting in the use of more efficient symmetric-key cryptography throughout. The SFCD handshake presented in \Cref{sec:protocol} allows for an arbitrary number of channels to be configured and created within the fleet, between drones and the GCS, as well as between drone pairs, with data authenticated and encrypted (where required) based on derived keys. The proposed SFCD packet format allows for up to 255 drones in a single mission, with a fleet supporting up to 256 missions at a time.

We demonstrated experimentally that our SFCD system achieves reasonable performance when used with low-cost consumer drones such as Parrot Bebop 2, showing only a 9.9\% reduction in flight time and bandwidths of at least 4.7Mibit/s at distances up to 260m---which may be further extended using the discussed multi-hop routing approach. This makes the proposed system suitable for applications involving real-time surveillance/monitoring of geographic areas spanning several square kilometres with a few tens of drones.

\section*{Acknowledgment}
The authors would like to thank Chris Bridges, Fran\c{c}ois Dupressoir and Alex Young for their work during initial discussions. Funded by UK Government.

\printbibliography

\end{document}